# Quantum transport evidence of the boundary states and Lifshitz transition in $Bi_4Br_4$

Dong-Yun Chen[1,2,3,4], Dashuai Ma[1,3,5], Junxi Duan[1,3], Dong Chen[4], Haiwen Liu[6], Junfeng Han[1,2,3*], and Yugui Yao[1,2,3]

[1]Centre for Quantum Physics, Key Laboratory of Advanced Optoelectronic Quantum Architecture and Measurement, Ministry of Education, School of Physics, Beijing Institute of Technology, Beijing 100081, China

[2]Yangtze Delta Region Academy of Beijing Institute of Technology, Jiaxing 314000, China

[3]Beijing Key Lab of Nanophotonics & Ultrafine Optoelectronic Systems, School of Physics, Beijing Institute of Technology, Beijing 100081, China

[4]College of Physics, Qingdao University, Qingdao 266071, China

[5]Institute for Structure and Function & Department of Physics, Chongqing University, Chongqing 400044, P. R. China

[6]Center for Advanced Quantum Studies, Department of Physics, Beijing Normal University, Beijing 100875, China

*Email: pkuhjf@bit.edu.cn

**The quasi-one-dimensional van der Waals compound $Bi_4Br_4$ was recently found to be a promising high-order topological insulator with exotic electronic states. In this paper, we study the electrical transport properties of $Bi_4Br_4$ bulk crystals. Two electron-type samples with different electron concentrations are investigated. Both samples have saturation resistivity behavior in low temperature. In the low-concentration sample, two-dimensional quantum oscillations are clearly observed in the magnetoresistance measurements, which are attributed to the band-bending-induced surface state on the (001) facet. In the high-concentration sample, the angular magnetoresistance exhibits two pairs of symmetrical sharp valleys with an angular difference close to the angle between the crystal planes (001) and (100). The additional valley can be explained by the contribution of the boundary states on the (100) facet. Besides, Hall measurements at low temperatures reveal an anomalous decrease of electron concentration with increasing temperature, which can be explained by the temperature-induced Lifshitz transition. These results shed light on the abundant surface and boundary state transport signals and the temperature-induced Lifshitz** transition in $Bi_4Br_4$.

**Introduction**

Since the discovery of topological insulators (TIs), they have been a hot topic in condensed matter physics and material science owing to their novel properties and potential applications [1-3]. In the band gap of the bulk state, TIs feature the gapless boundary states, which are protected by specific symmetries. $Z_2$ TIs, for example, are protected by the time-reversal symmetry [4,5]. Besides, with the protection of the point-group symmetries, the topological crystalline insulators (TCIs) and high-order topological insulators (HOTIs) are stabilized [6-8]. Different from $Z_2$ TIs and TCIs, the newly established HOTIs feature the one-dimensional (1D) gapless states in the hinges between crystal facets of a three-dimensional (3D) crystal, i.e., the hinge states. To date, there have been some predicted materials that are experimentally confirmed as the HOTIs by scanning-tunneling and photoemission spectroscopies [9-11]. However, there is still few reports on the transport properties of HOTIs, especially related to the boundary states.

Recently, the 1D van der Waals compound $Bi_4Br_4$ was predicted as a HOTI protected by $C_2$ rotation symmetry, and its hinge states have been observed by experiments [10-12]. The $Bi_4Br_4$ crystals are composed by the Bi-Br chains along *b* axis through van der Waals force in both other crystalline directions, as shown in the inset of Fig. 1(a) [13]. It has a semiconducting bulk structure with a relatively large band gap of about 0.2 eV [Fig. 1(b)]. The (001) monolayer of $Bi_4Br_4$ was identified as a quantum spin Hall insulator that hosts gapless helical edge states [14-16]. After stacking single layers into multilayers, the edge states of each monolayer are weakly coupled to each other and open a small gap, the (100) surface of $Bi_4Br_4$ bulk crystal is filled with coupled helical boundary states [17,18]. Owing to the special 1D crystal structure, both the (100) and (001) surfaces can be easily cleaved. The naturally cleaved (100) surface consists of many terraces and steps, making a number of hinges between (100) and (001) surfaces and topological hinge states are expected to be generated [11]. All the above properties make $Bi_4Br_4$ a promising platform to investigate the boundary states, i. e., the gapped edge states and the gapless hinge states.

In this paper, we report a transport study of $Bi_4Br_4$ single crystals. The results were measured on two respective samples, S1 and S2. In the magnetoresistance of S1, we observed SdH oscillations that only depend on the magnetic field perpendicular to (001) surface, which may originate from the band bending induced (001) surface state. By measuring the angular magnetoresistance (AMR), we observe the contributions which may attribute to the boundary states overall the (100) surface. With Hall measurements, the peculiar temperature dependence of the carrier concentrations in both samples are consistent with the scenario of the temperature-induced Lifshitz transition.

**Materials and methods**

High-quality single crystals of $Bi_4Br_4$ were grown by the self-flux method [19]. Strip-shaped crystals with the typical dimension of $2 \times 0.3 \times 0.1$ mm$^3$ were obtained, which can be easily cleaved along the longest dimension. The as-grown crystals were then characterized by x-ray diffraction in a PAN-analytical diffractometer with Cu-$K_\alpha$ radiation. The electric transport properties were measured using the five-probe method with current flowing along the long dimension. All the temperature and magnetic field dependent measurements were carried out in a Physical Properties Measurement System (Quantum Design, DynaCool-14) with temperature from 2 to 300 K and magnetic field up to 14 T. We used the Vienna ab initio simulation package (VASP) that is within the density functional theory (DFT) frame to perform the calculation of band structure [20]. Exchange correlation potential is treated within generalized gradient approximation (GGA) of Perdew-Burke-Ernzerhof (PBE) type [21]. In the calculation, we set the Monkhorst-Pack k-point mesh of $6 \times 6 \times 3$, and the energy cutoff of the plane-wave 300 eV is adopted. To obtain the Fermi surfaces for distinct energy, using the maximally localized Wannier functions (MLWF) method, we build a tight-binding model under the Wannier basis [22,23].

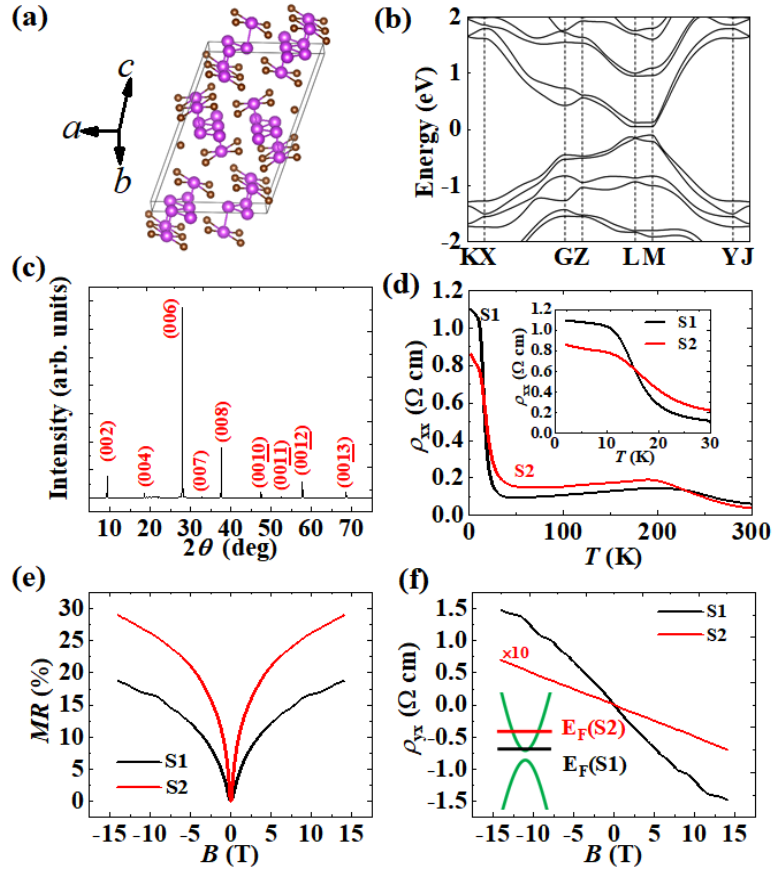

**Figure 1**. (a) The crystal structure of $Bi_4Br_4$, with the 1D Bi-Br chains along b axis as the building blocks. Bi and Br atoms are represented by the violet and brown balls, respectively. (b) The band structure of $Bi_4Br_4$. (c) The XRD pattern of $Bi_4Br_4$ measured on the maximum surface of a crystal. (d) The resistivity of samples S1 and S2 as the function of temperature. The inset shows the zoom-in view in low-temperature region. (e) The MR of the two samples measured at $T = 2$ K in a magnetic field normal to maximum surface of the crystal scanned from -14 to 14 T. Obvious SdH oscillations are present in the curve of S1. (f) Magnetic field dependence of Hall resistivities of the two samples measured at $T = 2$ K. The quite different Hall coefficients indicate the different positions of Fermi level, as sketched in the inset.

**Results and discussion**

Considering the 1D crystal structure of the $Bi_4Br_4$, the longest dimension of crystals can be determined as the *b* axis, while the *a* or *c* direction can be distinguished by measuring the XRD pattern of the maximum surface of a strip-shaped crystal, as presented in Figure 1(c). All the peaks can be identified as the (00*l*) diffractions of the $Bi_4Br_4$, indicating the maximum surface is the (001) surface. Figure 1(d) shows the temperature dependence of the resistivity for both samples S1 and S2 with the current along the $b$ axis. Similar to the previous reports, the resistivities of both samples exhibit a semiconductor behavior at high temperatures, they decrease with decreasing temperature in the intermediate region and increase again in low temperatures [19,24]. As shown in the inset in Fig. 1(d), our samples also show a trend of the resistance saturation below 10 K, which may indicate a special conducting channel in low temperatures, like the topological surface state in $SmB_6$ [25]. The obvious difference between S1 and S2 is the larger resistivity of S1 than that of S2 at low temperature. Another difference of these two samples is observed in magnetoresistance (MR) measurements with magnetic field perpendicular to (001) surface [Fig. 1(e)]. Compared to S2, the downward MR curve of S1 is superimposed with obvious magnetic oscillations which may benefit from the more insulating bulk state as discussed later. The downward sharp dip in the vicinity of zero field in both S1 and S2 is a typical weak anti-location (WAL) behavior [26,27]. The source of the resistivity and MR differences between two samples can be found from Hall measurements. As shown in Fig. 1(f), the slopes of the Hall resistivity curves of S1 and S2 are both negative but have quite different values, indicating the different concentrations of electrons. Using the one-band calculation, the electron concentration of S1 is only in the order of $10^{15}$ cm$^{-3}$, while the value in S2 is about two orders higher. The different electron concentrations indicate the different positions of Fermi level in the two samples, illustrated in the inset of Fig. 1(f). Besides, the magnetic field dependence of the Hall resistivity of S1 has a curved background except for the oscillating part, which can be well fitted by the two-band model, resulting in $n_1 = 1.1 \times 10^{15}$ cm$^{-3}$, $\mu_1 = 2484$ cm$^2$/Vs, $n_2 = 5.2 \times 10^{15}$ cm$^{-3}$, $\mu_2 = 573$ cm$^2$/Vs.

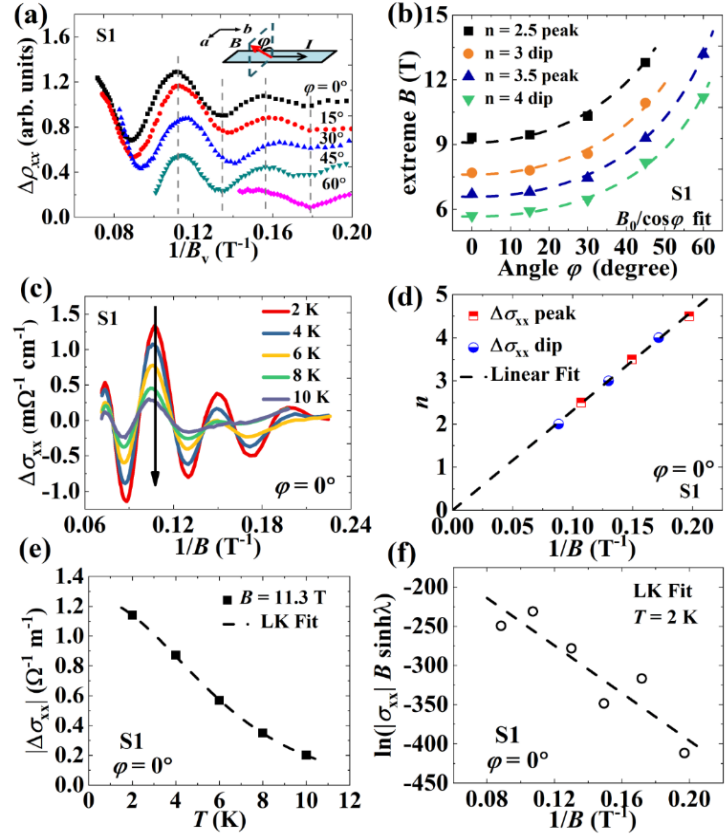

**Figure 2**. (a) Oscillation parts of resistivity measured under different field angles as the function of $1/B_v = 1/(B\cos(\varphi))$, where $\varphi$ is the angle between normal direction of the $ab$ plane and magnetic field rotated in the $ac$ plane. The inset illustrates the measurement geometry. (b) Angle dependence of the positions of various extremes in the SdH oscillations. A good fit to the 2D curves $B_0/\cos(\varphi)$ indicates the 2D nature of the oscillations. (c) The oscillations in the longitudinal conductivity $\sigma_{xx}$ with $\varphi = 0$ in different temperatures as the functions of the inversed field. The arrow shows the direction of temperature raising. (d) Landau index plot of the oscillations with the dips in the $\Delta\sigma_{xx}$ assigned to the integer numbers. (e) The LK fitting of the temperature dependence of oscillation amplitude $\Delta\sigma_{xx}$ at $B = 11.3$ T. (f) Dingle plot of the oscillations in $\sigma_{xx}$ at $T = 2$ K with the fitting line.

From Fig. 1(e) and 1(f), it should be noticed that both the MR and Hall curves of sample S1 have obvious quantum oscillations, which can give us more information of the corresponding electronic structure. Therefore, we furthermore measured its MR with the magnetic field rotated in the plane perpendicular to the current (the $ac$ plane). The measurement configuration is illustrated in the inset of Fig. 2(a), where $\varphi$ is the angle between the field and the normal direction of $ab$ plane. As shown in Fig. 2(a), the single-frequency quantum oscillations of S1 depend only on the field component perpendicular to the *ab* plane, suggesting a 2D behavior. This 2D nature can also be confirmed by the angle dependence of the extreme positions of the oscillations shown in Fig. 2(b). A natural speculation is that the QOs originate from the surface state on the (001) facet. However, according to the band structure calculations, the bulk Fermi surface of electron doped Bi$_4$Br$_4$ is also tube-like [Fig. 3(e)]. To distinguish the origin of the 2D QOs, we calculate the electron concentration from the Fermi surface area obtained from the QOs. From the Onsager relation that oscillating frequency $F = \frac{\hbar S_F}{2\pi e}$ is proportional to the cross section $S_F$ of contributing Fermi surface (FS), the $S_F$ at $\varphi = 0$ can be calculated as $2.20 \times 10^{-3} \frac{1}{\text{Å}^2}$, which gives a 2D carrier density $n_{2D}^{SdH} = \frac{k_F^2}{2\pi^2} = 1.12 \times 10^{12}$ cm$^{-2}$ with the spin degeneracy considered. Assuming the QOs come from the bulk state, the electron concentration can be calculated from the volume of FS $V_F = S_F c^*$ ($c^*$ is reciprocal lattice vector perpendicular to the *ab* plane) as $\frac{V_F}{4\pi^3} \approx 5.82 \times 10^{18}$ cm$^{-3}$. This value is two orders of magnitude larger than the value obtained from the Hall measurement, which can eliminate the bulk electric structure origin of the QOs. On the other hand, if we attribute the QOs to the surface state of the (001) facet and convert the 2D carrier density of surface state to the value in 3D form with the thickness $t$, the converted 3D carrier density $\frac{n_{2D}^{SdH}}{t} \approx 2.0 \times 10^{14}$ cm$^{-3}$ is close to the carrier density of the higher-mobility electron ($n_1 = 1.1 \times 10^{15}$ cm$^{-3}$) extracted from the two-band Hall fit. Therefore, the QOs are generated by the surface state of the (001) facet rather than the 2D FS the bulk state.

In SdH oscillations, the oscillatory part of longitudinal conductivity $\sigma_{xx}$ can be expressed in the Lifshitz-Kosevich (LK) theory: $\Delta\sigma_{xx} \sim \frac{2\pi^2 k_B T/\hbar\omega_c}{\sinh(2\pi^2 k_B T/\hbar\omega_c)} e^{-\lambda_D} \cos\left[2\pi\left(\frac{F}{B} + \frac{1}{2} - \gamma\right)\right]$, where $F$ is the oscillating frequency and $\gamma$ is related to Berry phase $\beta$ by $\beta = 2\pi\gamma$, $\lambda_D = 2\pi^2 k_B T_D/\hbar\omega_c$ with $T_D$ called the Dingle temperature and the cyclotron frequency $\omega_c = \frac{eB}{m_c}$ ($m_c$ is the cyclotron effective mass) [28]. The 0 Berry phase means a massive dispersion, while 1/2 corresponds to the massless one [29]. To get the Berry phase of the surface state, we employ the Landau index plot in Fig. 2(d), with the dip positions of $\Delta\sigma_{xx}$ corresponding to integer Landau indexes and the peak positions corresponding to half integer numbers [30]. The intercept of the linear fitting is around 0.03, giving a Berry phase close to zero. This is consistent with the theory that Bi$_4$Br$_4$ is a HOTI without topological surface state on (001) facet [11]. With the LK theory, we can further get the effective mass and mobility of the surface state from the QOs.

Figure 2(c) shows the oscillating conductivity $\Delta\sigma_{xx}$ at various temperatures, where the QOs decay with the raised temperature. By fitting the temperature dependence of $\Delta\sigma_{xx}$(11.3 T) with $\frac{2\pi^2 k_B T/\hbar\omega_c}{\sinh(2\pi^2 k_B T/\hbar\omega_c)}$ [Fig. 2(e)], the effective mass is estimated about 0.30 $m_e$. At a certain temperature, the amplitudes of oscillations decay by Dingle factor $\exp[-2\pi^2 k_B T_D/\hbar\omega_c]$ and $\Delta\sigma_{xx}$ vary with magnetic field as $\Delta\sigma \sim \frac{e^{-2\pi^2 k_B T_D/\hbar\omega_c}}{\sinh(2\pi^2 k_B T/\hbar\omega_c)}$. By Dingle analysis as shown in Figure 4(f), a Dingle temperature of 3.4 K can be obtained. With simple calculations, one can calculate the low limit mobility $\mu^{SdH}$ about 2072 cm$^2$/Vs, which is comparable with the value $\mu_1 = 2484$ cm$^2$/Vs obtained from the two-band Hall analysis. All the above results suggest the 2D SdH oscillations in S1 are from topological trivial surface state of (001) plane, which may stem from the band bending at the surface.

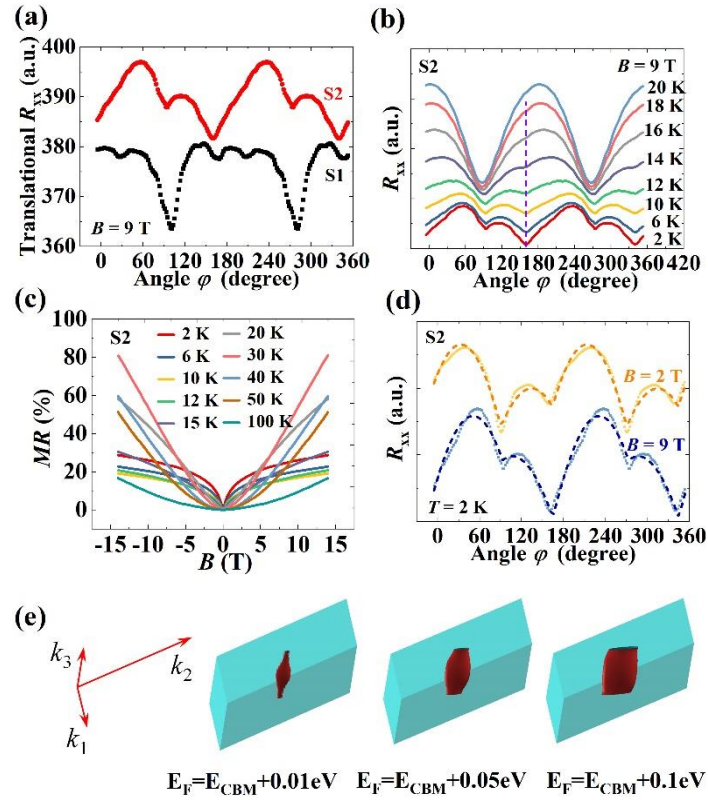

**Figure 3**. (a) AMR curves of samples S1 and S2 measured at $T$ = 2 K and $B$ = 9 T. (b) AMR curves of sample S2 measured under $B$ = 9 T at various temperatures. (c) MR curves of sample S2 measured at various temperatures. (d) AMR curves of sample S2 measured at $T$ = 2 K and $B$ = 2 and 9 T. The dashed lines are fitting curves of the contribution of two Fermi surfaces with a certain angle. (e) The Fermi surfaces of electron doped Bi$_4$Br$_4$ with Fermi levels located in different positions. All of Fermi surfaces have a tube-like shape.

AMR measurement is a powerful tool to analyze the shape of fermi surface, which can enhance our understanding of the electron structure of $Bi_4Br_4$. We carried out the measurement on both samples over a $\varphi$ range of -5° to 355°. Figure 3(a) shows the comparison of the AMR curves of the two samples measured at $T$ = 2 K and $B$ = 9 T. The AMR curve of sample S1 has a pair of symmetrical sharp valleys and several shallow valleys, the sharp valley has an angle difference of around 90° compared with a maximum value. This phenomenon may be roughly explained by the 2D AMR of the bulk electron structure of $Bi_4Br_4$ and the superposed fluctuations caused by the QOs. While for sample S2, the AMR curve is different, which exhibits two pairs of symmetric sharp valleys and wide maxima, where the angle difference of the two nearby dips is around 70° or 110°. Unaffected by quantum oscillations, the AMR of the sample S2 is cleaner and suitable for further analysis. To gain insight into the peculiar AMR behavior, we also measured the AMR and MR curves of sample S2 at various temperatures as shown in Fig. 3(b) and 3(c), respectively. In the AMR curves, the minimum at about 160° becomes weakened as temperature increases, and eventually disappears at 20 K. The MR curves change from the downward curves at low temperatures to the parabolic behavior above 20 K. Both of AMR curves and MR curves have a crossover behavior from low temperatures to high temperatures. The low-temperature AMR behaviors of sample S2 are hard to be explained only by the bulk electron structure of $Bi_4Br_4$. To elucidate this, we calculate the Fermi surfaces of the electron doped $Bi_4Br_4$ with the Fermi levels locating at different positions which are deep enough compared to the carrier density of S2, as shown in Fig. 3(e). The tube-like Fermi Surfaces indicate the quasi-2D band structure of $Bi_4Br_4$ in the *ab* plane, showing no signature of special AMR symmetry as S2 in low temperatures. The additional minimum in the AMR of sample S2 at low temperatures suggests another possible contribution beside the 2D bulk state. Combining the position difference of the minima is close to the crystal constant β = 107.42°, we assume that this is caused by the surface state on the (100) facet. To further verify this hypothesis, we fit the AMR curves of sample S2 measured under 2 T and 9 T at 2 K by the expression: $a|\cos(\varphi - \varphi_0)| + |\cos(\varphi - \varphi_0 + \Delta\varphi)| + c$ where $\Delta\varphi$ is the angle difference between the two components. As shown in Fig. 3(d), the AMR curves can be well fitted, and the fitted $\Delta\varphi$ values are around 106.2° and 108.2°. It suggests that the first term comes from the 2D bulk state, while the second term may come from the contribution of the surface state of (100) facet. This scenario is also in line with the temperature dependent AMR and the shortcut characteristics in resistivity: as the temperature increases, the contribution of the surface state gradually decreases and the bulk state is eventually dominant at high temperatures. That surface state may be composed of a number of hinge states and gapped edge states in the (100) facet.

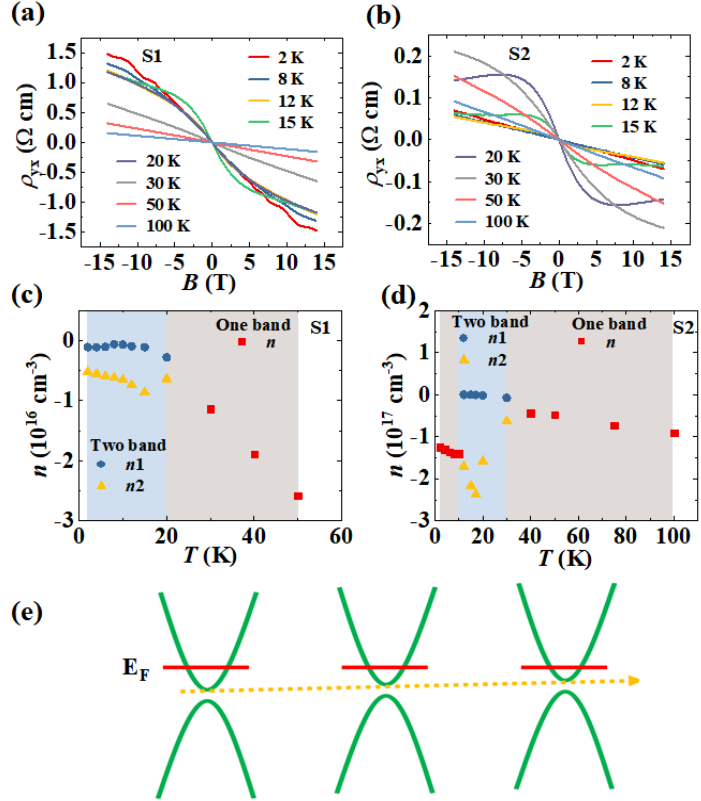

**Figure 4**. (a, b) Magnetic field dependence of Hall resistivity of sample S1 and S2 at various temperatures, respectively. (c, d) Temperature dependent carrier concentrations of samples S1 and S2, respectively. (e) A sketch for the temperature induced Lifshitz transition. The band structure shift upward with temperature raising.

Comparing with S1 and S2 above, the energy positions of Fermi level make great effect on the transport properties of $Bi_4Br_4$. To investigate the evolution of carrier densities, we measured the magnetic field dependence of Hall resistivity of both samples S1 and S2 at different temperatures, as presented in Fig. 4(a) and (b). At high temperatures, both samples show a linear field dependence of Hall resistivity with a negative slope. As temperature decreases, Hall resistivity curves of both samples become bended below 20K, which suggest the contributions of more than one band and can be fitted by two-band model. Figure 4(c) and (d) show the carrier concentrations of the two samples that have been obtained by one-band calculation and two-band fitting [31]. At temperatures below 20 K, there are two electron-type carriers in both samples, one of which has a significantly higher density than the other one. Considering a single Fermi surface in the electron doped $Bi_4Br_4$, the minority carrier should originate from the surface state on the (001) facet that is discovered above. By distributing the concentration of the surface electrons into the whole crystal of S1, we find that it is comparable with the smaller concentration. Moreover, it's very interesting that there is a carrier density decline with temperature rising in a low temperature region, especially in sample S2. This unconventional phenomenon is contrary to the conventional thermal excitation theory [32], but can be explained by the temperature-induced Lifshitz transition recently discovered by ARPES experiment in $Bi_4Br_4$ [33]. As sketched in Fig. 4(e), the band structure shifts upward relative to the Fermi energy with the temperature increased, leading to a reduced electron concentration at higher temperatures. The complicated temperature dependence of the carrier concentration in low temperatures may be the consequence of the competition between the thermal excitation and the temperature induced Lifshitz transition. Due to the different band filling, samples S1 and S2 show different electrical properties dominated by different electrons. The lower Fermi level for sample S1 results in the higher bulk resistivity and the larger ratio of the surface conductivity contribution. The lower Fermi level also gives rise to a smaller effective mass of the surface state, which improves the mobility. Thus, sample S1 can exhibit obvious SdH oscillations of surface state on the (001) surface. On the other hand, the higher Fermi level in sample S2 leads to the higher conductivity for both bulk and surface/boundary states, so that the conductivity of boundary states on the (100) surface can contribute obvious signals in the AMR curves. Furthermore, it makes sense that anomalous carrier evolution with temperature is more pronounced in sample S2 with a higher Fermi level if we attribute it to the temperature induced Lifshitz transition.

**Conclusion**

In summary, we carry out a comprehensive study of the transport properties of $Bi_4Br_4$ single crystals. MR and Hall resistivity of two representative samples, S1 and S2, are investigated, which are both electron-type with different electron concentrations. Sample S1 exhibits obvious 2D SdH oscillations in the MR curves, which we attribute to the trivial surface state on the (001) facet. Sample S2 has two pairs of symmetrical sharp valleys and wide maxima in the low-temperature AMR curves, which may originate from the contribution of the boundary states on the (100) facet with a great number of terraces and steps. The Hall resistivity of sample S2 shows an abnormally higher electron concentrations at low temperatures than those in a high temperature range. This unconventional temperature dependence may be explained by the temperature induced Lifshitz transition. Our results reveal the possible transport signals of the topological boundary states and provide transport evidence of the temperature induced Lifshitz transition in $Bi_4Br_4$.

**Acknowledgement**

We thank the valuable discussions and selfless help from Shuang Jia. This work is funded by the National Natural Science Foundation of China (NSFC) (11734003), the National Key R&D Program of China (2020YFA0308800), the Strategic Priority Research Program of Chinese Academy of Sciences (XDB30000000) and the Beijing Natural Science Foundation (Grant No. Z210006).